\documentclass[preprint,showpacs,12pt,floatfix,aps]{revtex4}
\usepackage{graphicx}
\usepackage{latexsym}
\usepackage{amsmath}

\newcommand{\myfigure}[3]{
        \begin{figure}
        \centerline{
        \includegraphics{#1.eps}}
        \caption{#2}
        \label{#3}
        \end{figure}
}

\newcommand{\bq}{\begin{equation}}
\newcommand{\eq}{\end{equation}}
\newcommand{\bqn}{\begin{eqnarray}}
\newcommand{\eqn}{\end{eqnarray}}
\newcommand{\nb}{\nonumber}
\newcommand{\lb}{\label}
\newcommand{\rr}{\bf r}

\begin{document}
\title{How the Charge Can Affect the Formation of Gravastars}
\author{R. Chan $^{1}$}
\email{chan@on.br}
\author{M.F.A. da Silva $^{2}$}
\email{mfasnic@gmail.com}
\affiliation{$^{1}$ Coordena\c{c}\~ao de Astronomia e Astrof\'{\i}sica, 
Observat\'orio
Nacional, Rua General Jos\'e Cristino, 77, S\~ao Crist\'ov\~ao, CEP 20921-400,
Rio de Janeiro, RJ, Brazil\\
$^{2}$ Departamento de F\'{\i}sica Te\'orica,
Instituto de F\'{\i}sica, Universidade do Estado do Rio de Janeiro,
Rua S\~ao Francisco Xavier 524, Maracan\~a,
CEP 20550-900, Rio de Janeiro - RJ, Brasil} 
\date{\today}

\begin{abstract}
In recent work we physically interpreted a special gravastar solution
characterized by a zero Schwarzschild mass. In fact,  in that case, none 
gravastar was formed and the shell expanded, leaving behind a de Sitter or a 
Minkowski spacetime, or collapsed without forming an event horizon, originating 
what we called a massive non-gravitational object. This object has two 
components of non zero mass but the exterior spacetime is Minkowski or de Sitter.
One of the component is a massive thin shell and the other one is de Sitter spacetime
inside. The total mass of this object is zero Schwarzschild mass, 
which characterizes an exterior vacuum spacetime.
Here, we extend this study
to the case where we have a charged shell. Now, the exterior is a Reissner-Nordstr\"om 
spacetime and, depending on the parameter $\omega=1-\gamma$ of the equation of state
of the shell, and the charge, a gravastar structure can be formed. We have found that 
the presence of the charge contributes to the stability of the gravastar, if the
charge is greater than a critical value.  Otherwise, a massive non-gravitational 
object is formed for small charges.
\end{abstract}


\maketitle

\section{Introduction}

As alternatives to black holes,  gravastars have received some attention
recently \cite{grava1}\cite{grava2}, partially due to the tight connection between the
cosmological constant and a currently accelerating universe \cite{DEs},
although very strict observational constraints on the existence of such
stars may exist \cite{BN07}.

The pioneer model of gravastar was proposed by Mazur and Mottola (MM) \cite{MM01}.  After this work,
Visser and Wiltshire (VW) \cite{VW04} reduced
the number of shells of the original model from five to three. The interior
was described by the de Sitter metric, while the exterior by the Schwarzschild
metric. Between both there is a thin shell constituted by stiff fluid, which
is located in a such way that it is outside of the Schwarzschild event
horizon and inside of the de Sitter one, in order to eliminate both horizons
from the whole spacetime. They also pointed out that there are two different
types of stable gravastars which are stable gravastars and "bounded excursion" gravastars.
The first one represents a stable structure already formed, while the second one
is a system with a shell which oscillates around a equilibrium position which can loose energy and
to stabilize at the end.

Recently we have done an extensive study on the problem of the stability of gravastars.
The first model \cite{JCAP} consisted of an internal de Sitter spacetime, 
a dynamical infinitely thin  shell of
stiff fluid, and an external Schwarzschild spacetime, as proposed by VW \cite{VW04}.
We have shown explicitly that the final output can be a black
hole, a "bounded excursion" stable gravastar, a Minkowski, or a de Sitter spacetime,
depending on the total mass $m$ of the system,  the cosmological constant $\Lambda$, and
the initial position $R_{0}$ of the dynamical shell.
Therefore, we have shown, for the first time in the literature, that although it does
exist a region of the space of the initial parameters where it is always formed
stable gravastars, it still exists a large region of this space where
we can find black hole formation.  Then, we conclude that gravastar is not
an alternative model to black hole as it was originally proposed by VW models \cite{VW04}.

In the second paper \cite{JCAP1}, we have generalized the previous work on the problem of
stable gravastars considering an equation of state $p = (1-\gamma)\sigma$ for
the shell, instead of only using a stiff fluid ($\gamma=0$).
We have found that stable gravastars can be formed even for $\gamma \ne 0$,
since $\gamma < 1$, generalizing the gravastar models proposed until now.
We also have confirmed the previous results, i.e., that both gravastars and black
holes can be formed, depending on the initial parameters.

In the third work \cite{JCAP2}, we have generalized the former one considering now
an interior constituted by an anisotropic dark energy fluid.
We have again confirmed the previous results, i.e., that both gravastars and black
holes can be formed, depending on the initial parameters.
It is remarkable that for this case we have an interior filled by a physical matter,
instead of a de Sitter vacuum.  Thus, it is similar to phantom energy star models.

In the fourth paper \cite{JCAP3}, we generalized a previous model of gravastar 
consisted of an internal de Sitter spacetime, a dynamical infinitely thin  shell with
an equation of state, but we have considered an external
de Sitter-Schwarzschild spacetime. We have found that the exterior cosmological 
constant imposes a limit
to the gravastar formation, i.e., the exterior cosmological constant must be
smaller than the interior cosmological constant.

In another work \cite{CQG2010}, 
we investigated a particular solution, which emerges from the 
gravastar studies, 
with zero Schwarzschild mass, which implies in a non-gravitational 
object. This kind of structure can be  possible, since the 
gravitational mass depends on 
the trace of the energy momentum tensor, instead of the energy density only. 
As the inner region is filled by dark energy, there is an equilibrium 
between the repulsive gravitational effect (due to the inner region) and the attractive 
one (due to the thin shell) on all the test particles in the exterior, 
vanishing the gravitational interactions. So, the object that we have studied is similar to a 
point-like topological defect, since the vacuum solutions do not describe all the spacetime.

Before the present work,  Carter \cite{Carter} studied spherically symmetric gravastar 
solutions which possess an
(anti) de Sitter interior and a Reissner-Nordstr\"om exterior. 
He followed the same approach that Visser and
Wiltshire took in their work \cite{VW04} assuming a potential $V(a)$ and
then finding the equation of state of the shell. He have found a wide range
of parameters which allows stable gravastar solutions, and presented the
different qualitative behaviors of the equation of state for these
parameters.  Gravastar with charge has been also studied by others authors \cite{Horvat}.

It is the aim of this work to generalize that solution for zero Schwarzschild 
mass, but now, considering a charged shell, in order to investigate the effect 
of the charge on the stability of the gravastar formation.

The paper is organized as follows.  In Section II we present the charged gravastar model.
In Section III a discussion on the definition of the gravitational mass of the charged gravastar is done.
In Section IV we find the potential of the gravastar's shell and we study two special 
cases, that are: i) dust fluid ($\gamma=1$) and ii) stiff fluid ($\gamma=0$).  
Finally, in Section V we present the final remarks.

\section{Charged Gravastar Model}

The interior spacetime is described by the de Sitter metric given by 
\bq
ds^2_{i}=-fdt^2 + \frac{1}{f}dr^2 + r^2 d\Omega^2,
\lb{ds2-}
\eq
where $f=1- (r/L_i)^2$, $L_i=\sqrt{3/\Lambda_i}$ and 
$d\Omega^2 = d\theta^2 + \sin^2(\theta)d\phi^2$.

The exterior spacetime is given by a Reissner-Nordstr\"om metric, which 
characterizes a vacuum charged exterior spacetime 
\bq
ds^2_{e}= - h dv^2 + \frac{1}{h} d{\rr}^2 + {\rr}^2 d\Omega^2,
\lb{ds2+}
\eq
where $h=1 - \frac{2m}{\rr}+ (\frac{q}{\rr})^2 $.

In order to follow the pioneer model of gravastar proposed by Mazur and Mottola (MM) \cite{MM01} and
Visser and Wiltshire (VW) \cite{VW04}, we have assumed an inner spacetime with cosmological constant. 
The presence of a cosmological constant only in the inner spacetime can be understood in the context 
where it emerges from a phase transition of the quantum vacuum near or at the place where the event 
horizon is expected to form. Thus, it does not need to be present in the whole spacetime.
 
The metric of the hypersurface do the shell is given by
\bq
ds^2_{\Sigma}= -d\tau^2 + R^2(\tau) d\Omega^2.
\lb{ds2Sigma}
\eq

Since $ds^2_{i} = ds^2_{e} = ds^2_{\Sigma}$ then $r_{\Sigma}={\rr}_{\Sigma}=R$,
and besides
\bq
\dot t^2=\left[ f - \frac{1}{f}\left( \frac{\dot R}{\dot t} \right)^2 \right]^{-1}= \left[ 1 - \left(\frac{R}{L_i}\right)^2 \right]^{-2} \left[1- \left(\frac{R}{L_i}\right)^2 + \dot{R}^2 \right] ,
\lb{dott2}
\eq
and
\bq
\dot v^2=\left[ h - h^{-1} \left( \frac{\dot R}{\dot v} \right)^2 \right]^{-1}= \left[1 - \frac{2m}{\rr}+ \left(\frac{q}{\rr} \right)^2\right]^{-2} \left[ \dot{R}^2 + 1 -2 \frac{m}{R} + \left(\frac{q}{R}\right)^2 \right]  ,
\lb{dotv2}
\eq
where the dot represents the differentiation with respect to $\tau$.

Thus, the interior and exterior normal vector are given by
\bq
n^{i}_{\alpha} = (-\dot R, \dot t, 0 , 0 ),\qquad n^{e}_{\alpha} = (-\dot R, \dot v, 0 , 0 ).
\lb{nalpha-}
\eq

The interior and exterior extrinsic curvature are given by
\bqn
K^{i}_{\tau\tau}&=& -[(3 L_i^4 \dot R^2 - L_i^4 \dot t^2 + 2 L_i^2 R^2 \dot t^2 -
R^4 \dot t^2) R \dot t -(L_i+R) (L_i-R) (\dot R \ddot t - \ddot R \dot t) L_i^4] \times \nb \\
& &(L_i+R)^{-1} (L_i-R)^{-1} L_i^{-4}
\lb{Ktautau-}
\eqn
\bq
K^{i}_{\theta\theta}=\dot t (L_i+R) (L_i-R) L_i^{-2} R
\lb{Kthetatheta-}
\eq
\bq
K^{i}_{\phi\phi}=K^{i}_{\theta\theta}\sin^2(\theta),
\lb{Kphiphi-}
\eq
\bqn
K^{e}_{\tau\tau} & = &
- \left\{ \left[ (2 m R \dot v - q^2  \dot v + R^2  \dot R - R^2  \dot v )
(2 m R \dot v - q^2  \dot v - R^2  \dot R - R^2  \dot v) - R^4  \dot R^2
- R^4  \dot R^2 \right] \right. \times \nb \\
& & \left. (m R - q^2 ) \dot v
- (q^2  + R^2  - 2 m R) (\dot R \ddot v - \ddot R \dot v) R^5 \right\}
\left(q^2  + R^2  - 2 m R \right)^{-1}   R^{-5},
\lb{Ktautau+}
\eqn
\bq
K^{e}_{\theta\theta}=\dot v \left(q^2 +R^2 - 2 m R \right) R^{-1},
\lb{Kthetatheta+}
\eq
\bq
K^{e}_{\phi\phi}=K^{e}_{\theta\theta}\sin^2(\theta).
\lb{Kphiphi+}
\eq

Since we have \cite{Lake}
\bq
[K_{\theta\theta}]= K^{e}_{\theta\theta}-K^{i}_{\theta\theta} = - M,
\lb{M}
\eq
where $M$ is the mass of the shell, thus
\bq
M = \dot{t} (L_i + R)(L_i  - R) \frac{R}{L_i^2} - \dot{v}  (q^2 + R^2 - 2m R)
 \frac{1}{R}.
\lb{M1}
\eq

Then, substituting equations (\ref{dott2}) and (\ref{dotv2}) into (\ref{M1}) 
we get
\bq
M - R \left[ 1 -\left(\frac{R}{L_i}\right)^2  + \dot R^2 \right]^{1/2}
+ 
R  \left[1-\frac{2m}{R} + \dot R^2  + \left(\frac{q}{R}\right)^2\right]^{1/2}  
=0.
\lb{M2}
\eq

Solving the equation (\ref{M2}) for $\dot R^2/2$ we obtain the potential 
$V(R,m,L_i,q)$.
In order to keep the ideas of our work \cite{JCAP1} as much as possible, we 
consider the thin shell as consisting
of a fluid with a equation of state, $\vartheta = (1-\gamma)\sigma$, where 
$\sigma$ and $\vartheta$ denote, 
respectively, the surface energy density and pressure of the shell and $\gamma$
is a constant.  The equation of motion of the shell is given by \cite{Lake}
\bq
\dot M + 8\pi R \dot R \vartheta = 4 \pi R^2 [T_{\alpha\beta}u^{\alpha}n^{\beta}]=
\pi R^2 \left(T^e_{\alpha\beta}u_e^{\alpha}n_e^{\beta}-T^i_{\alpha\beta}u_i^{\alpha}n_i^{\beta} \right),
\lb{dotM}
\eq
where $u^{\alpha}$ is the four-velocity and $n^{\beta}$ is its orthogonal 
vector.  Since the interior region is constituted by a vacuum with cosmological
constant and the exterior correspond to a vacuum with an electromagnetic field, 
characterized by the Reissner-Nordstr\"om spacetime, it is easy to show that
\bq
\dot M + 8\pi R \dot R (1-\gamma)\sigma = 0,
\lb{dotM1}
\eq
since \cite{Adler} 
\bq
T_{\alpha\beta}^e = F_{\alpha\lambda}F^{\lambda}_{\beta}+\frac{1}{4}F_{\lambda\nu}F^{\lambda\nu}= 
\frac{q^2}{2\rr^4}\left( h\delta^{v}_{\alpha}\delta^{v}_{\beta}-
h^{-1}\delta^{\rr}_{\alpha}\delta^{\rr}_{\beta}+
\rr^2\delta^{\theta}_{\alpha}\delta^{\theta}_{\beta}+
\rr^2 \sin^2\theta \delta^{\phi}_{\alpha}\delta^{\phi}_{\beta} \right),
\eq
where $F_{\alpha\lambda}$ is the Maxwell tensor.

Since $\sigma = M/(4\pi R^2)$  we can solve equation (\ref{dotM1}) giving
\bq
M=k R^{2(\gamma-1)},
\lb{Mk}
\eq
where $k$ is an integration constant.

Substituting equation (\ref{Mk}) into $V(R,m,L_i,q)$, we obtain the general 
expression for the potential, 
\bqn
\lb{potential}
V(R,m,L_i,q,\gamma) & = & \frac{1}{2} -  \frac{1}{4}{\frac {{R}^{2}}{{L_i}^{2}}} 
- \frac{1}{8}{\frac {{R}^{10}}{
{R}^{4\gamma} {L_i}^{4}}} + \frac{1}{4}{\frac {{q}^{2}}{
{R}^{2}}}- \frac{1}{2}{\frac {{m}}{R}} + \frac{1}{2}{\frac {{R}^{3}{m}\,{q}
^{2}}{ {R}^{4\gamma} }}  \nb \\
 &-&  \frac{1}{8}{\frac { {R}^{4\gamma}}{{R}^{6}}}- \frac{1}{4}{\frac {{R}
^{6}{q}^{2}}{ {R}^{4\gamma} {L_i}^{2}}}-  \frac{1}{8}
\frac {{R}^{2}{q}^{4}}{{ {R}^{4\gamma} }} \nb \\
&+& \frac{1}{2}{\frac {{R}^{7} m}{ {R}^{4\gamma}
{L_i}^{2}}} - \frac{1}{2}{\frac {
{R}^{4}{m}^{2}}{ {R}^{4\gamma}}}
\eqn
where we have redefined the Schwarzschild mass $m$, the cosmological constants
$L_i$, the charge $q$ and the radius $R$ as $m \equiv m k^{-\frac{1}{2\gamma-3}}$,
$L_i \equiv L_i k^{\frac{2}{2\gamma-3}} $,
$q \equiv q k^{\frac{2}{2\gamma-3}}$,
$R \equiv R k^{-\frac{1}{2\gamma-3}}$.

\section{Total Gravitational Mass}

In order to study the gravitational effect generated by the two components of 
the
gravastar, i.e., the interior de Sitter and the thin shell in the exterior 
region, we need to calculate the total gravitational mass of a spherical symmetric 
system. Some alternative definitions are given by \cite{Marder},\cite{Israel} and \cite{Levi}. 
Here we consider the Tolman formula for the mass, given by 
\bq
M_{G}=\int_0^{R_0} \int_{-\pi}^{\pi} \int_0^{2\pi} \sqrt{-g}\; 
T^\alpha_\alpha dr d\theta d\phi,
\eq
where $\sqrt{-g}$ is the determinant of the metric.  For the special case of a 
thin shell we have
\bq
M_{G}=\int_0^{R_0} \int_{-\pi}^{\pi} \int_0^{2\pi} \sqrt{-g}\; 
T^\alpha_\alpha \delta({\rr}-R_0) d{\rr} d\theta d\phi.
\eq
Since the energy-momentum tensor of the shell is given by
\bq
T_{\alpha\beta} = T^{(F)}_{\alpha\beta}+T^{(EM)}_{\alpha\beta},
\eq
where the ($F$) denotes the fluid and ($EM$) the electromagnetic tensor and 
\bq
T^{(EM)}_{\alpha\lambda}g^{\lambda\beta}=
\frac{q^2}{2{\rr}^4}\left(-\delta^{v}_{\alpha}\delta_{v}^{\beta}-
\delta^{{\rr}}_{\alpha}\delta_{{\rr}}^{\beta}+
\delta^{\theta}_{\alpha}\delta_{\theta}^{\beta}+
\delta^{\phi}_{\alpha}\delta_{\phi}^{\beta} \right).
\eq
Thus, the gravitational mass of the thin shell is given by
\bq
M_G^{shell}=(3-2\gamma)M,
\eq
since $T^{(EM)}_{\alpha\beta}g^{\alpha\beta}=0$.

For the interior de Sitter (dS) spacetime we have
\bq
M_{G}^{(dS)}=-\frac{2}{3} \Lambda_i R_0^3.
\eq
Then, the de Sitter
interior presents a negative gravitational mass, since $\Lambda_i > 0$, 
in accordance to the its repulsive effect.

Now we can write the total gravitational mass of the gravastar as
\bq
\label{MGtotal}
M_G^{(total)}=M_G^{(shell)}+M_G^{(dS)}= (3-2\gamma)M-\frac{2}{3}\Lambda_i R_0^3.
\eq
This mass also represents the Schwarzschild exterior mass ($m=M_G^{(total)}$) 
for this gravastar.
Thus, it is possible to obtain a physical structure through a combination of 
these two solutions,
which results in a system with $m=0$, a Reissner-Nordstr\"om exterior spacetime,
without mass but with charge.

\section{The Potential for $m=0$}

Firstly, we can see that in this case, the exterior metric has no event horizon. 
So, the initial radius of the shell must be in the interval $0\leq R_0\leq L_i$.
For the particular case where the Schwarzschild mass is vanished, $m=0$, 
the potential (\ref{potential}) is written as
\bqn
V(R,L_i,q,\gamma) & = & \frac{1}{2}-\frac{1}{4}\,{\frac {{R}^{2}}{{L_i}^{2}}}-\frac{1}{4}\,{\frac {{q}^{2}{R}^{6}
}{{L_i}^{2} {R}^{4\gamma} }}+\frac{1}{4}\,{\frac {{q}^{
2}}{{R}^{2}}}-\nb \\
& & \frac{1}{8}\,{\frac { {R}^{4\gamma} }{{R}^{6}}
}-\frac{1}{8}\,{\frac {{R}^{10}}{  {R}^{4\gamma}{L_i}^{
4}}}-\frac{1}{8}\,{\frac {{R}^{2}{q}^{4}}{ {R}^{4\gamma} }} ,
\lb{VR1}
\eqn
and differentiating the potential we get
\bqn
\frac{dV(R)}{dR} & = & -\frac{1}{2}\,{\frac {R}{{L_i}^{2}}}-\frac{3}{2}\,{\frac {{q}^{2}{R}^{5}}{{L_i}^{2} {R}^{4\gamma} }}-\frac{1}{2}\,{\frac {{q}^{2}}{{R}^{3}}}+\frac{3}{4}\,{\frac { {R}^{4\gamma} }{{R}^{7}}}-\frac{5}{4}\,{\frac {{R}^{9}}{ {R}^{4\gamma} {L_i}^{4}}}\nb \\
& & -\frac{1}{4}\,{\frac {R{q}^{4}}{ {R}^{4\gamma} }}+{\frac {{R}^{5}\gamma\,{q}^{2}}{{L_i}^{2} {R}^{4\gamma} }}-\frac{1}{2}\,{\frac {\gamma\, {R}^{4\gamma} }{{R}^{7}}}+\frac{1}{2}\,{\frac{{R}^{9}\gamma}{ {R}^{4\gamma} {L_i}^{4}}}+\frac{1}{2}\,{\frac {R\gamma\,{q}^{4}}{ {R}^{4\gamma}}}.
\lb{V2}
\eqn

A second differentiation of the potential furnishes
\bqn
\frac{d^2V(R)}{dR^2} & = & -\frac{1}{2L_i^2}+11\,{\frac {{R}^{4}\gamma\,{q}^{2}}{{L_i}^{
2} {R}^{4\gamma} }}-\frac{15}{2}\,{\frac {{q}^{2}{R}^{4}}{{
L_i}^{2}  {R}^{4\gamma} }}+\frac{3}{2}\,{\frac {{q}^{2}}{
{R}^{4}}}+\nb \\
& & \frac{13}{2}\,{\frac {\gamma\, {R}^{4\gamma} }{{R}
^{8}}}+\frac{19}{2}\,{\frac {{R}^{8}\gamma}{ {R}^{4\gamma}{
L_i}^{4}}}-{\frac {45}{4}}\,{\frac {{R}^{8}}{ {R}^{4\gamma}
 {L_i}^{4}}}+\frac{3}{2}\,{\frac {\gamma\,{q}^{4}}{ {R
}^{4\gamma} }}-\nb \\
& & -\frac{1}{4}\,{\frac {{q}^{4}}{ {R}^{4\gamma}
 }}-4\,{\frac {{q}^{2}{R}^{4}{\gamma}^{2}}{{L_i}^{2}
 {R}^{4\gamma}}}-2\,{\frac { {R}^{4\gamma}
 {\gamma}^{2}}{{R}^{8}}}-\nb \\
& & 2\,{\frac {{R}^{8}{\gamma}^{2}}{
 {R}^{4\gamma} {L_i}^{4}}}-2\,{\frac {{q}^{4}{
\gamma}^{2}}{ {R}^{4\gamma} }}-{\frac {21}{4}}\,{
\frac { {R}^{4\gamma} }{{R}^{8}}}. 
\lb{V3}
\eqn
In the following we will consider some particular equations of state, 
representing some different kinds of energy, shown in the figure \ref{fig10}. 

\subsection{Stiff Fluid Shell  ($\gamma=0$):}

\myfigure{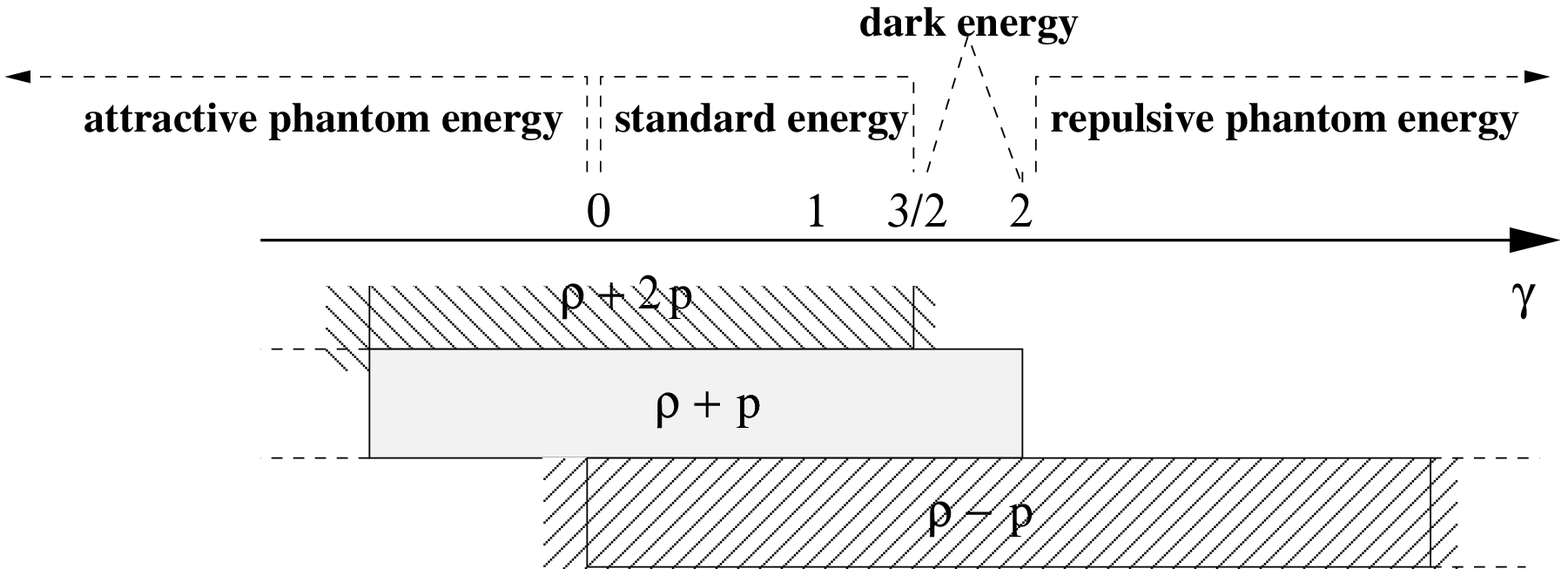}{This is a plot with the different 
kinds of the shell's matter depending on the parameter $\gamma$.}
{fig10}

In order to simplify, firstly we have considered a shell constituted by stiff fluid, 
as it was done in the original gravastar�s model, meaning that we choose 
$\gamma=0$. Vanishing the first derivative of the potential gives us the 
critical values of the parameter $L_i$, or the positive values of ${L_i}^2$ 
which assure that the potential has a profile as shown in the figure \ref{fig1}. 

\myfigure{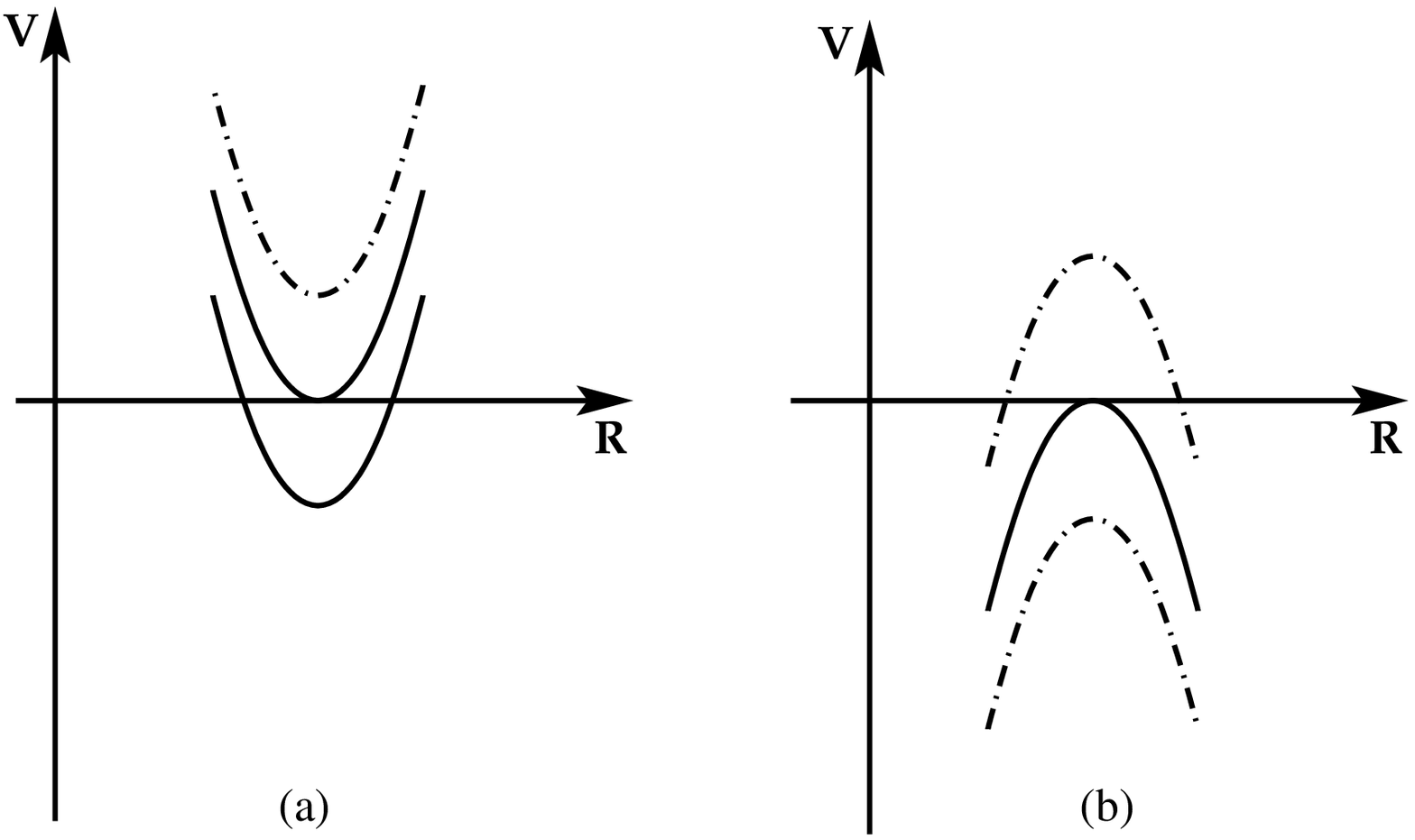}{This plot shows the possible potentials.}
{fig1}

Doing this we obtain
\bq
L_{ic}^2 = -\frac {1}{\Delta_1}\left( {3{R}^{12}{q}^{2}} +{{R}^{8}}+
{2{R}^{8}\Delta_2} \right),
\lb{Lic}
\eq
where $\Delta_1=2\,{q}^{2}{R}^{4}-3+{R}^{8}{q}^{4}$ and  
$\Delta_2= \sqrt {{R}^{8}{q}^{4}-{q}^{2}{R}^{4}+4}$.
Substituting $L_i^2$ by $L_{ic}^2$ into the second derivative of the potential, 
we obtain the expression
\bqn
\label{der2pot}
& &\frac{d^2V(R)}{dR^2} = \nb \\
& &\frac{1}{\Delta_3} \left( -\frac {192}{{R}^{8}}
+8{R}^{8}{q}^{8}-96{q}^{4}-\frac{24\Delta_2}{{R}^{8}}+
48q^4\Delta_2 -\frac{100 q^2\Delta_2}{{R}^{4}} + 12{R}^{4}{q}^{6}\Delta_2
+28{R}^{4}{q}^{6}+ \frac {124{q}^{2}}{{R}^{4} } \right), \nb \\
\eqn
where $\Delta_3=\left( 3{q}^{2}{R}^{4}+1+2\Delta_2 \right)^2$.

\myfigure{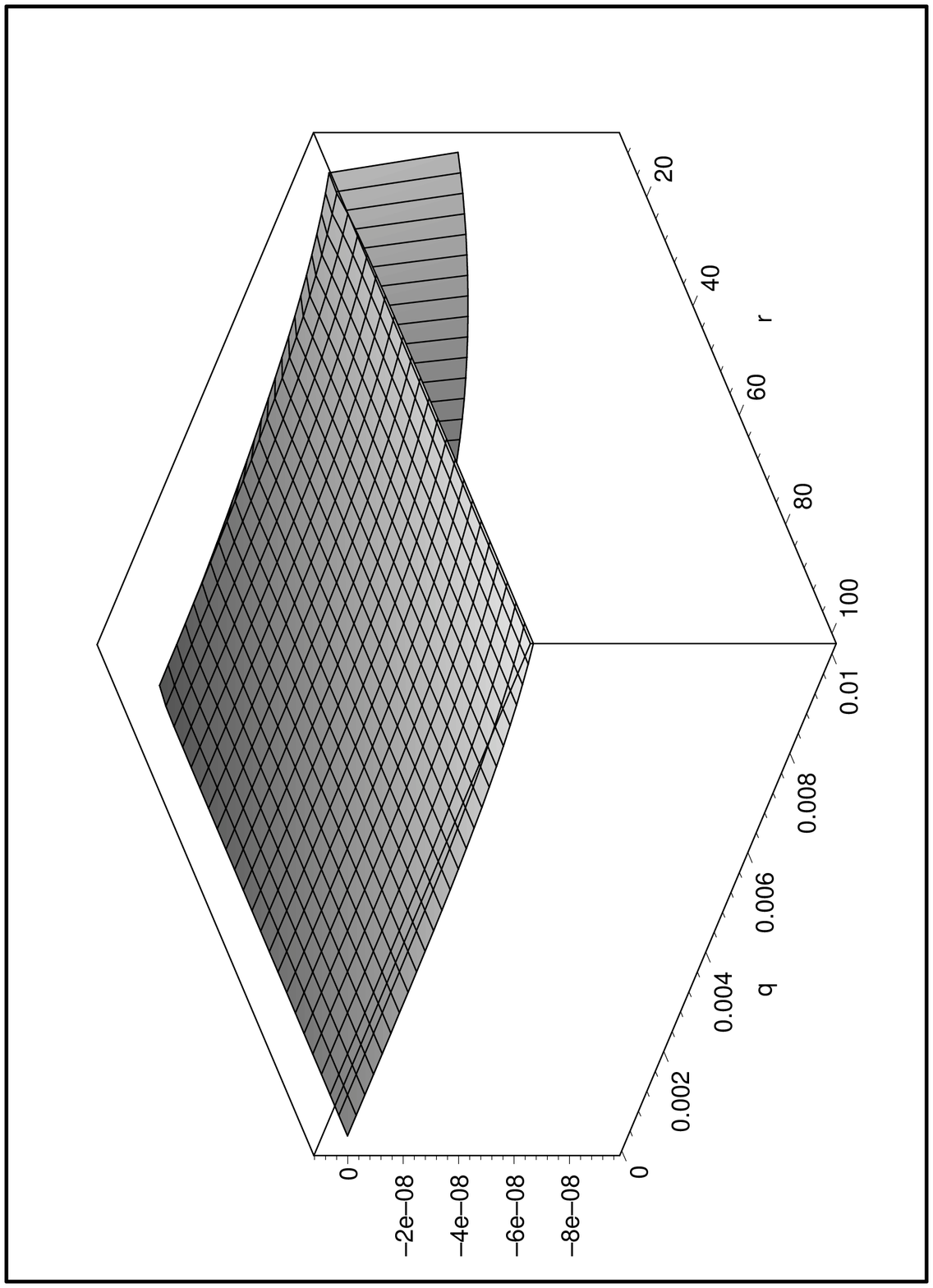}{This plot shows, in terms of $R$ and $q$, the second 
derivative of the potential $V(R,q)$, for $\gamma=0$ and for the 
intervals $0 \le q \le 0.01$ and $15 \le R \le 100$.}
{fig2}

\myfigure{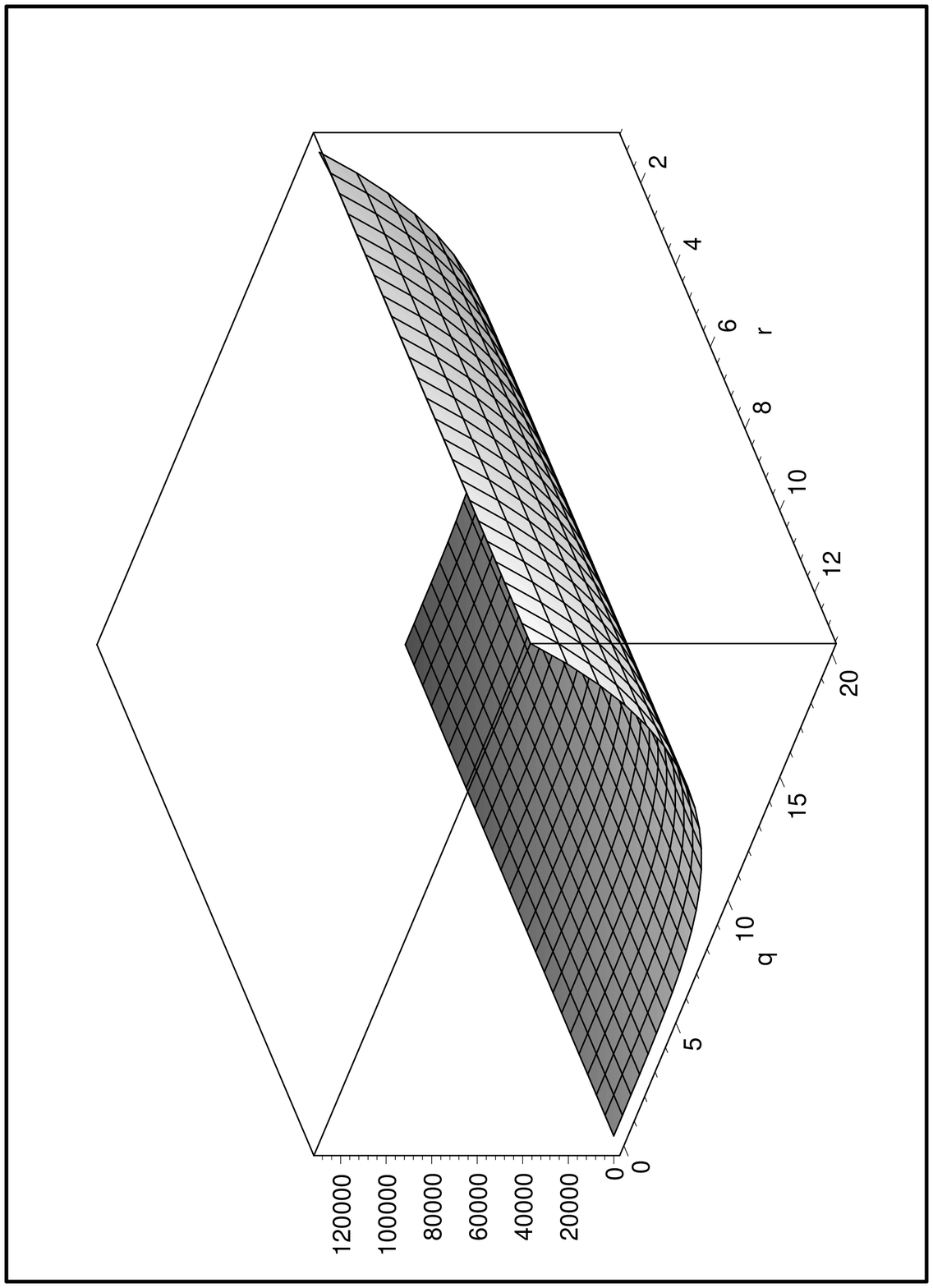}{This plot shows, in terms of $R$ and $q$, the second 
derivative of the potential $V(R,q)$, for $\gamma=0$ and for the 
intervals $0 \le q \le 20$ and $0 \le R \le 15$.}
{fig3}

Then we can see, from figures \ref{fig2} and \ref{fig3}, that the potential can 
present a minimum value.
From figure \ref{fig4} we can see that it 
represents a stable configuration (or 
oscillating) since its minimum value is not positive. 

\myfigure{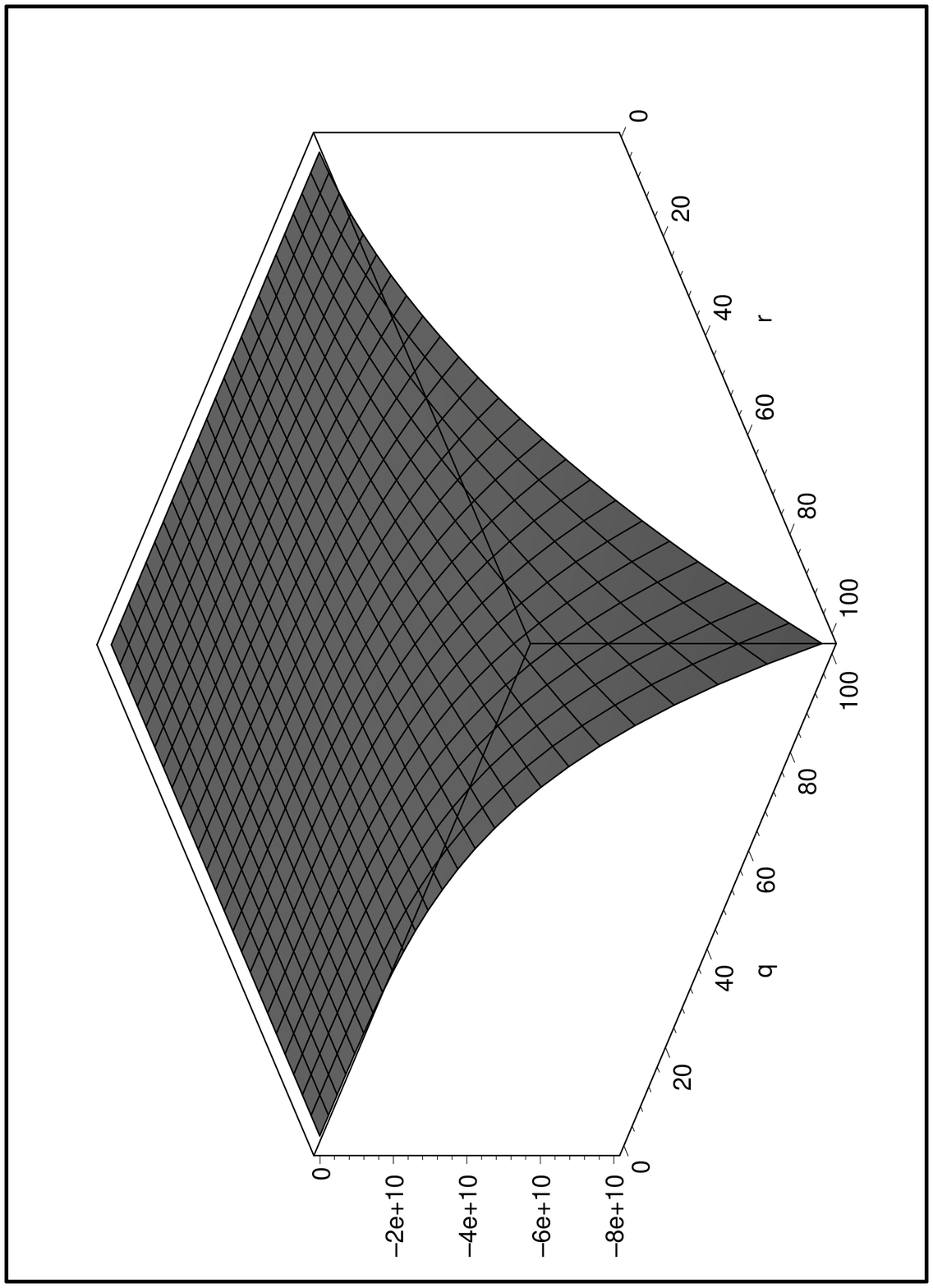}{This plot shows, in terms of $R$ and $q$,
the  minimum of the potential $V(R,q)$, for 
$\gamma=0$ for the intervals $0 \le q \le 100$ and $0 \le R \le 100$. }
{fig4}

Substituting $L_i$ by $L_{ic}$ into the potential, and equating it to zero, we 
obtain the critical values of $q$, in order to guarantee that the maximum 
of the potential is zero, which is given by
\bq
q_c = \frac{1}{\sqrt {6}}\sqrt {{\frac {{A}^{2/3}+92\,{R}^{6}+52+4\,{R}^
{12}-2\,\sqrt [3]{A}{R}^{6}+10\,\sqrt [3]{A}}{{R}^{4}\sqrt [3]{A}}}},
\lb{Rc}
\eq
where  
$$ A=399\,{R}^{12}+1092\,{R}^{6}+280-8\,{R}^{18}+3\,\sqrt {3}\sqrt {-
 \left( 16\,{R}^{18}+21\,{R}^{12}+48\,{R}^{6}+16 \right)  \left( -12+5
\,{R}^{6} \right) ^{2}}.
$$

Note that there is only one possibility to have a real solution for 
$q_c=1.918822876$ , corresponding to the radius $R_c=\sqrt[6]{\frac{12}{5}}=
1.157093730$. Thus, for a stiff fluid shell there is an unique pair of the 
parameters $R_c$ and $q_c$ which is able to form a stable gravastar. 
The graphics of $V'$ and $V"$ stablish a minimum limit to the charge 
in order to have gravastar formation, which is reinforced by the 
absence of gravastar with zero mass and zero charge in previous paper \cite{JCAP2}. 
Then, we can conclude that $q\geq q_c$ to form gravastar, where $q=q_c$ 
represents a stable gravastar, while $q>q_c$ represents bounded excursion gravastars.

These results allow us to conclude that the presence of charge in the shell 
created the necessary conditions to form a stable gravastar even with a zero 
Schwarzschild mass.

\subsection{Dust Fluid Shell ($\gamma=1$):}

Now,  we consider one shell constituted by dust fluid, meaning that we choose 
$\gamma=1$.
Again, vanishing of the first derivative of the potential gives us the critical
values of the parameter $L_i$ .  Doing this we obtain
\bq
L_{ic}^2 =\frac {1}{\Delta_5} \left( R^{4}{q}^{2}+{R}^{4}+2{R}^{4}\Delta_4 \right),
\lb{Lic1}
\eq
where $\Delta_4=\sqrt {{q}^{4}-{q}^{2}+1}$ and $\Delta_5=-2\,{q}^{2}+1+{q}^{4}$.
Putting this into the second derivative of the potential, we  obtain the 
expression
\bqn
\label{der2pot1}
& &\frac{d^2V(R)}{dR^2} = \nb \\
& & \frac{1}{R^4\Delta_6} \left(-8 +24q^2-8q^8+24q^6-32q^4-4q^6\Delta_4+4q^2\Delta_4+4q^4\Delta_4-4\Delta_4 \right),
\eqn
where $\Delta_6=\left( {q}^{2}+1+2\Delta_4 \right)^2$.

\myfigure{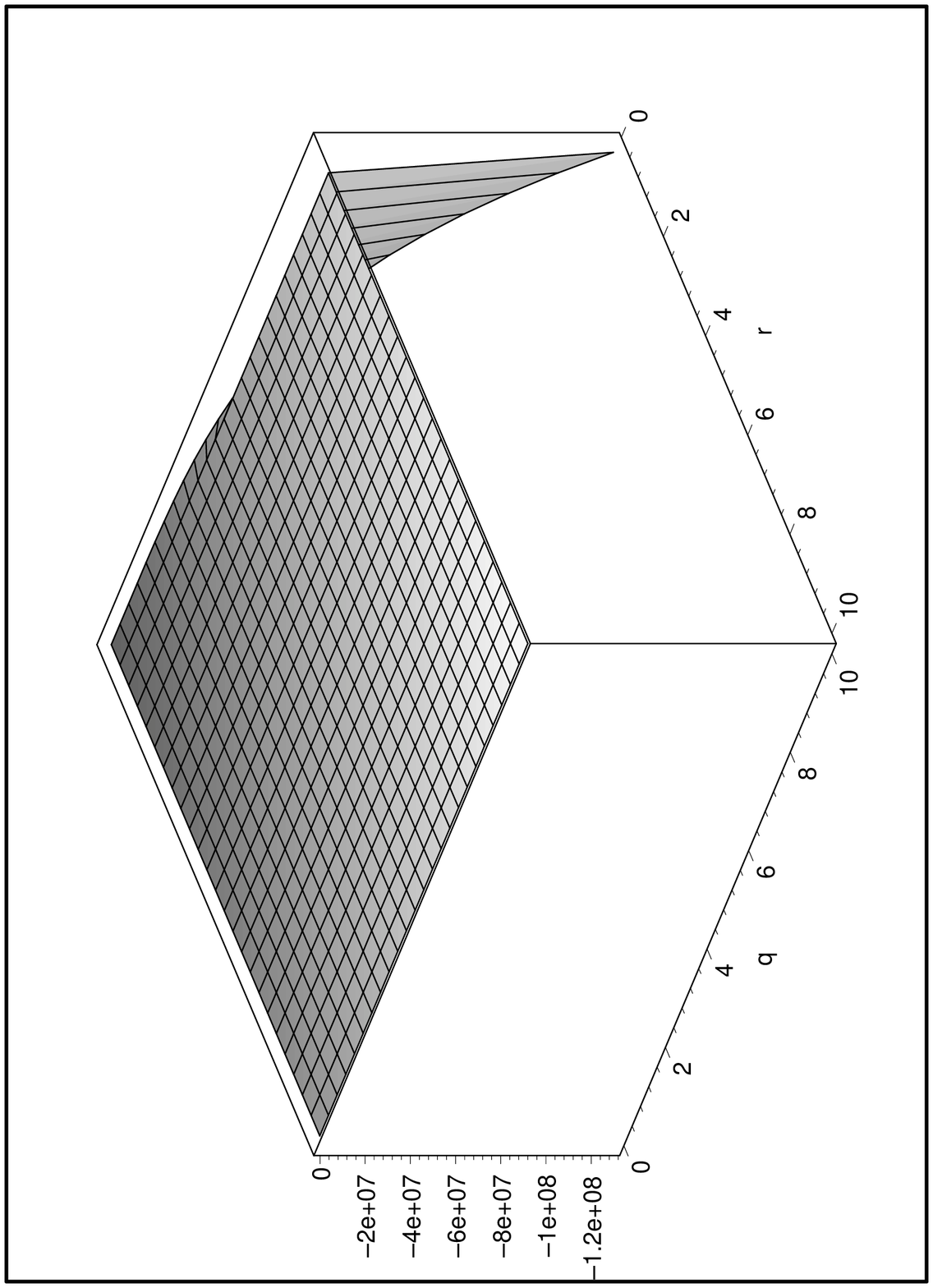}{This plot shows, in terms of $R$ and $q$, the second 
derivative of the potential $V(R,q)$, for $\gamma=1$ and for the 
intervals $0 \le q \le 10$ and $0 \le R \le 10$.}
{fig5}

\myfigure{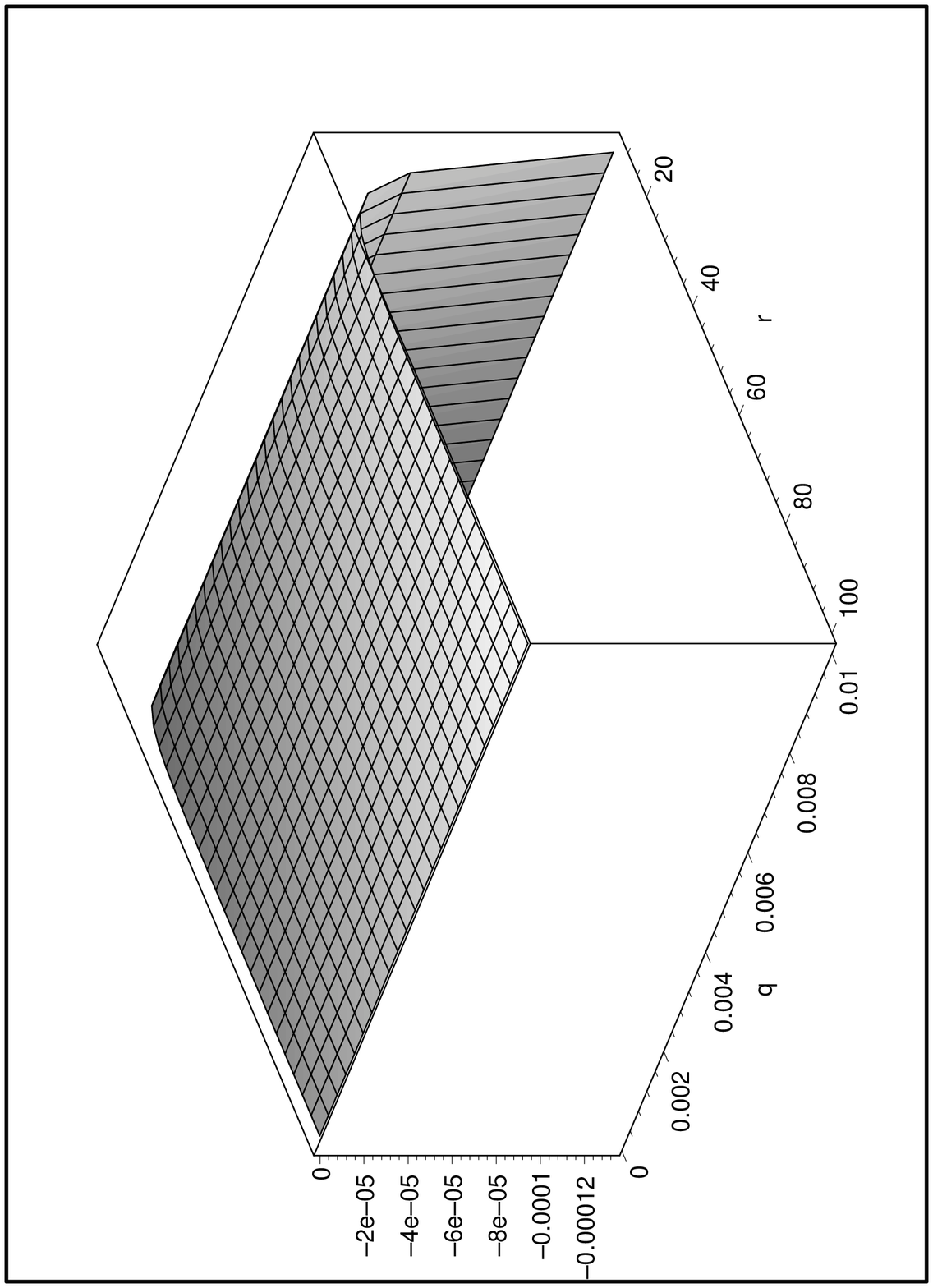}{This plot shows, in terms of $R$ and $q$, the second 
derivative of the potential $V(R,q)$, for $\gamma=1$ and for the 
intervals $0 \le q \le 0.01$ and $10 \le R \le 100$.}
{fig6}

Then we can see, from figures \ref{fig5} and \ref{fig6} , that the second 
derivative of the potential 
is always negative. Thus, the potential of a charged dust shell cannot 
represent a stable (or even a bounded excursion) gravastar 
configuration, independently of the charge of the shell.

\section{Final Remarks}

In this paper, we have generalized the problem of the stability of gravastars
studied recently by us \cite{JCAP2}, introducing a charged shell
with a Reissner-Nordstr\"om exterior spacetime. Thus, the model consists of a de Sitter
interior spacetime, a dynamical infinitely thin shell of
fluid with an equation of state $p=(1-\gamma)\sigma$, and an external
Schwarzschild spacetime.

In order to simplify the mathematical calculus and to isolate the effect
of the charge in the gravastar formation,
we investigated a particular solution which emerges from the
gravastar studies with zero Schwarzschild mass, which implies in a non-gravitational
object.

For the stiff fluid for the shell ($\gamma=0$), we have shown that for $q\geq q_c$ it forms gravastar: 
when $q=q_c$ it forms a stable gravastar, while $q>q_c$ it forms bounded excursion gravastar.

For the dust shell ($\gamma=1$) the potential of a charged dust shell cannot
represent a stable (or even a bounded excursion) gravastar
configuration, independently of the charge of the shell.

In summary, we have found that
the presence of the charge contributes to the stability of the gravastar, if the
charge is greater than a critical value.  Otherwise, a massive non-gravitational
object is formed for small charges.

\begin{acknowledgments}
The financial assistance from 
FAPERJ/UERJ (MFAdaS) is gratefully acknowledged. The
author (RC) acknowledges the financial support from FAPERJ (no.
E-26/171.754/2000, E-26/171.533/2002 and E-26/170.951/2006). 
MFAdaS and RC also acknowledge the financial support from 
Conselho Nacional de Desenvolvimento Cient\'{\i}fico e Tecnol\'ogico - 
CNPq - Brazil.  The author (MFAdaS) also acknowledges the financial support
from Financiadora de Estudos e Projetos - FINEP - Brazil (Ref. 2399/03).
\end{acknowledgments}

\end{document}